\documentclass[pra,aps,graphicx,multicol,epsfig]{revtex4}
\usepackage{graphics}

\begin{document}

\title{An effective potential for one-dimensional matter-wave solitons in an
axially inhomogeneous trap}
\author{Sergio De Nicola$^{1}$, Boris A. Malomed$^{2}$, Renato Fedele$^{3}$}
\affiliation{$^{1}$Istituto di Cibernetica " E Caianiello " del CNR, Comprensorio ``A.
Olivetti", Via Campi Flegrei, 34, I-80078 Pozzuoli (Na), Italy }
\affiliation{$^{2}$Department of Interdisciplinary Studies, School of Electrical
Engineering, Faculty of Engineering, Tel Aviv University, Tel Aviv 69978,
Israel}
\affiliation{$^{3}$Dipartimento di Scienze Fisiche, Universit\'{a} Federico II and INFN
Sezione di Napoli, Complesso Universitario di M.S. Angelo, via Cintia,
I-80126, Napoli, Italy}

\begin{abstract}
We demonstrate that a tight transverse trap with the local frequency, $%
\omega _{\perp }$, gradually varying in the longitudinal direction ($x$)
induces an effective potential for one-dimensional solitons in a
self-attractive Bose-Einstein condensate. An analytical approximation for
this potential is derived by means of a variational method. In the lowest
approximation, the potential is $N(S+1)\omega _{\perp }(x)$, with $N$ the
soliton's norm (number of atoms), and $S$ its intrinsic vorticity (if any).
The results can be used to devise nonuniform traps helping to control the
longitudinal dynamics of the solitons. Numerical verification of the
analytical predictions will be presented elsewhere.

\medskip\medskip

\begin{center}
{ To be published in Physics Letters A}
\end{center}

\end{abstract}

\maketitle

\section{Introduction}

A Bose-Einstein condensate (BEC) with attractive interactions between atoms
(negative scattering length) may be stable in an external trap if the number
of atoms in the condensate is below a collapse threshold \cite{Pethik}. In
this case, the BEC in a nearly one-dimensional (1D) \textquotedblleft
cigar-shaped" trap, which features tight confinement in the transverse plane
and a weak potential along the longitudinal axis, can form stable
matter-wave packets in the form of bright solitons. A single soliton \cite%
{Khaykovich02} and multi-soliton complexes \cite{Strecker02} were created in
the $^{7}$Li condensate loaded into a strongly elongated optical trap. More
recently, solitons whose shape is nearly three-dimensional (3D), were
observed in a post-collapse state in a condensate of $^{85}$Rb atoms \cite%
{Cornish} (as well as in $^{7}$Li, in this experiment the sign of the
interatomic interactions was switched into attractive by means of the
Feshbach resonance).

A fundamental equation which provides for a very accurate description of the
dynamics of a rarefied quantum gas of Boson atoms in the mean-field
approximation is the 3D Gross-Pitaevskii (GP) equation \cite{Pethik}. In a
normalized form, the equation is%
\begin{equation}
i\frac{\partial \psi }{\partial t}=-\frac{1}{2}\left( \nabla _{\perp }^{2}+
\frac{\partial ^{2}}{\partial x^{2}}\right) \psi +\left[ \frac{1}{2}\omega
_{\perp }^{2}\left( y^{2}+z^{2}\right) +U(x)\right] \psi +g|\psi |^{2}\psi ,
\label{GPE}
\end{equation}%
where $\psi $ is the single-atom wave function, $x$ and $y,z$ are the
longitudinal (axial) and transverse coordinates (the transverse Laplacian $%
\nabla _{\perp }^{2}$ acts on $y$ and $z$), $\omega _{\perp }$ is the
frequency accounting for the tight transverse confinement, $U(x)$ a loose
axial potential, and $g$ a scaled nonlinearity constant (in the case of
self-attraction, $g$ is negative).

In the experiment, the transverse trapping potential may be axially
nonuniform, which corresponds to $\omega _{\perp }=\omega _{\perp }(x)$ in
Eq. (\ref{GPE}). It may also depend on time, hence $\omega _{\perp }=\omega
_{\perp }(x,t)$, in the most general case. Actually, the axial nonuniformity
is an unavoidable feature of any experimental setup, and, on the other hand,
specially designed nonuniformity may be used as an additional tool for the
control of dynamics of trapped solitons. The objective of this Letter is to
derive an effective longitudinal potential induced (in addition to the
explicitly present potential, $U(x)$ in Eq. (\ref{GPE})) by the $x$%
-dependence of $\omega _{\perp }$, and the corresponding equation of motion
for axial solitons.

If the transverse potential is much stronger than the longitudinal one, it
is natural to reduce the full 3D GP equation, (\ref{GPE}), to an effective
1D equation. Different approaches were proposed to achieve this purpose \cite%
{Shlyap}-\cite{Brand} under different conditions. In most cases, the
reduction is based on assuming a factorized \textit{ansatz} for the wave
function \cite{Shlyap,Luca},
\begin{equation}
\psi (x,y,z,t)=\exp \left( -\frac{r^{2}}{2\sigma ^{2}(x,t)}\right) ~\frac{%
f(x,t)}{\sqrt{\pi }\sigma (x,t)},  \label{factorizing}
\end{equation}%
where $r^{2}\equiv y^{2}+z^{2}$, the transverse width $\sigma $ and local
amplitude $f$ being slowly varying functions of $x$ and $t$. Using the
Lagrangian representation of Eq. (\ref{GPE}), effective 1D equations for $%
\sigma $ and $f$ can be derived as variational equations. First, $\sigma $
is eliminated in favor of $f$,
\begin{equation}
\sigma ^{2}=\omega _{\perp }^{-1}\sqrt{1+\left( g/2\pi \omega _{\perp
}\right) \left\vert f\right\vert ^{2}},  \label{sigma^2}
\end{equation}%
and then a closed-form 1D GP equation with \emph{nonpolynomial nonlinearity}
is derived for $f$ \cite{Luca}:
\begin{equation}
i\frac{\partial f}{\partial t}=-\frac{1}{2}\frac{\partial ^{2}f}{\partial
x^{2}}+\left[ \omega _{\perp }\frac{1+\left( 3g/4\pi \omega _{\perp }\right)
|f|^{2}}{\sqrt{1+\left( g/2\pi \omega _{\perp }\right) \left\vert
f\right\vert ^{2}}}+U(x)\right] f~.  \label{final}
\end{equation}%
Equation (\ref{final}) with $g<0$ admits stable solitary-wave solutions,
which were analyzed in Ref. \cite{Luca} too.

In the case of the weak nonlinearity, $\left( g/2\pi \omega _{\perp }\right)
\left\vert f\right\vert ^{2}\ll 1$, the nonlinear term in Eq. (\ref{final})
may be expanded in powers of $|f|^{2}$, which leads to the cubic GP equation
with an additional quintic term that corresponds to higher-order
self-attraction,%
\begin{equation}
i\frac{\partial f}{\partial t}=-\frac{1}{2}\frac{\partial ^{2}f}{\partial
x^{2}}+\omega _{\perp }\left( \frac{g}{2\pi \omega _{\perp }}\left\vert
f\right\vert ^{2}-\frac{3g^{2}}{8\pi ^{2}\omega _{\perp }^{2}}|f|^{4}\right)
f+\left[ \omega _{\perp }+U(x)\right] f~.  \label{CQ}
\end{equation}%
In a more direct form, a 1D equation with the \textit{cubic-quintic} (CQ)\
nonlinearity of this type was derived and employed in Refs. \cite%
{Shlyap,Brand}. A formally similar equation of the CQ type (but with a
self-defocusing quintic term) was considered earlier in works aiming to take
into regard three-body collisions in BEC \cite{Fatkhulla}). Despite the
possibility of collapse induced by the self-focusing quintic term in the 1D
setting, the CQ equation has a family of exact soliton solutions, which are
\emph{stable} against small perturbations \cite{Lev} (these new solutions
were obtained as an analytical continuation of well-known soliton solutions
\cite{Pushkarov} to the CQ equation with the self-defocusing quintic term).

The approach based on the factorized ansatz (\ref{factorizing}) was
generalized in Ref. \cite{Luca2} to describe configurations with intrinsic
vorticity, with the ansatz replaced by%
\begin{equation}
\psi (x,y,z,t)=r^{S}\exp \left( -\frac{r^{2}}{2\sigma ^{2}(x,t)}+iS\theta
\right) ~f_{S}(x,t),  \label{S}
\end{equation}%
where $\theta $ is the angular coordinate in the $(y,z)$ plane, and $%
S=1,2,...$ is the integer vorticity (which naturally carries with itself the
pre-exponential factor $r^{S}$). Actually, $S$ plays the role of the
\textquotedblleft spin" of effectively 1D solitons generated by the ansatz.
This way, an equation for $f_{S}(x,t)$ similar to Eq. (\ref{final}) can be
derived.

The Letter is organized as follows. In the next section, using the
variational approximation \cite{Progress}, we develop a framework for the
analysis of the soliton's dynamics in the model with the axially nonuniform
trapping, $\omega _{\perp }=\omega _{\perp }(x)$ (the variational approach
employs the Hamiltonian, rather than Lagrangian). In Section 3, we derive a
final result, \textit{viz}., an effective axial potential for the soliton
induced by the $x$-dependence of $\omega _{\perp }$, and the corresponding
equation of motion for the soliton. The analytical results are obtained
under the natural assumption that a scale of the variation of $\omega
_{\perp }(x)$ is much larger than the size of the soliton. The paper is
concluded by Section 4.

\section{Variational analysis}

Aiming to derive an effective potential and equation of motion for solitons
in the case of $\omega _{\perp }=\omega _{\perp }(x)$, we skip the
derivation of an effective 1D equation, and instead adopt a 3D ansatz for
the soliton which, in the general case, includes the intrinsic vorticity
(cf. Eq. (\ref{S})):
\begin{equation}
\psi _{\mathrm{sol}}=Ar^{S}\exp \left( -\frac{r^{2}}{2\sigma ^{2}}+iS\theta
\right) \mathrm{sech}\left( \frac{x-\xi }{W}\right) e^{i\phi },
\label{ansatz}
\end{equation}%
with amplitude $A$, longitudinal width $W$, central coordinate $\xi $ and
phase $\phi $, in addition to the transverse width, $\sigma $, and spin, $%
S=0,1,2,...$, that were defined above. The norm of this ansatz is%
\begin{equation}
N\equiv 2\pi \int_{0}^{\infty }rdr\int_{-\infty }^{+\infty }\left\vert \psi
_{\mathrm{sol}}(r,x)\right\vert ^{2}=2\pi S!A^{2}W\sigma ^{2(S+1)}~.
\label{N}
\end{equation}

To derive an effective potential for the soliton as a function of the
coordinate $\xi $, we use the Hamiltonian of three-dimensional equation (\ref%
{GPE}),
\begin{equation}
H=\pi \int_{-\infty }^{+\infty }dx\int_{0}^{\infty }rdr\left[ \left\vert
\psi _{x}\right\vert ^{2}+\left\vert \psi _{r}\right\vert ^{2}+\left( \omega
_{\perp }^{2}(x)r^{2}+2U(x)\right) |\psi |^{2}+g|\psi |^{4}\right] .
\label{H}
\end{equation}%
The substitution of ansatz (\ref{ansatz}) in $H$ and straightforward
integrations yield%
\begin{eqnarray}
\frac{H}{\pi } &=&(1+S)!A^{2}W\sigma ^{2S}+\frac{S!}{3}\frac{A^{2}}{W}\sigma
^{2(1+S)}+(1+S)!\omega _{\perp }^{2}(\xi )A^{2}W\sigma ^{2(2+S)}  \nonumber
\\
&&+2S!U(\xi )A^{2}W\sigma ^{2(1+S)}+\frac{g}{3}\frac{(2S)!}{2^{2S}}%
A^{4}W\sigma ^{2(1+2S)}.  \label{Hnext}
\end{eqnarray}%
This result was obtained under the condition that, as said above, $\omega
_{\perp }(x)$ varies on a scale which is much longer than the soliton's
width, $W$, the same being assumed about the axial potential, $U(x)$. For
this reason, $\omega _{\perp }^{2}(x)$ and $U(x)$ in Hamiltonian (\ref{Hnext}%
) are taken at $x=\xi $.

The soliton's peak density, $A^{2}$, can be eliminated, in Eq. (\ref{Hnext}%
), in favor of the conserved norm, $N$, using Eq. (\ref{N}). This leads to
\begin{eqnarray}
H &=&\left( 1+S\right) \frac{N}{2}\left[ \frac{1}{\sigma ^{2}(\xi )}+\omega
_{\perp }^{2}(\xi )\sigma ^{2}(\xi )\right] +\frac{1}{6}\frac{N}{W^{2}}
+NU(\xi )+\frac{C_{S}}{3\pi }\frac{gN^{2}}{W\sigma ^{2}}~,  \label{HW} \\
C_{S} &\equiv &\frac{(2S)!}{2^{2(1+S)}(S!)^{2}}~.  \label{C}
\end{eqnarray}
In this expression, the transverse and longitudinal widths of the soliton, $%
\sigma $ and $W$, are free parameters. The soliton chooses their values by
minimizing the Hamiltonian, which leads to conditions
\begin{equation}
\frac{\partial H}{\partial W}=\frac{\partial H}{\partial \left( \sigma
^{2}\right) }=0.  \label{opt}
\end{equation}
The first condition takes a simple form, making it possible to eliminate the
longitudinal width,
\begin{equation}
W=-\frac{\pi }{C_{S}}\frac{\sigma ^{2}}{gN},  \label{W}
\end{equation}%
which, obviously, makes sense only for $g<0$. Indeed, bright solitons may
only exist in the BEC with the attractive nonlinearity, corresponding to
negative $g$ (unless a periodic optical-lattice potential is present, that
can support \textit{gap solitons} at $g>0$ \cite{GS,HS}). The substitution
of $W$ from expression (\ref{W}) in Eq. (\ref{HW}) yields%
\begin{equation}
H=\left( 1+S\right) \frac{N}{2}\left[ \frac{1}{\sigma ^{2}(\xi )}+\omega
_{\perp }^{2}(\xi )\sigma ^{2}(\xi )\right] +NU(\xi )-\frac{C_{s}^{2}}{6\pi
^{2}}\frac{g^{2}N^{3}}{\sigma ^{4}}~.  \label{Hsigma}
\end{equation}

Further, the second condition from Eq. (\ref{opt}) amounts to a cubic
equation for $\sigma ^{-2}$:%
\begin{equation}
\frac{2C_{s}^{2}}{3\pi ^{2}\left( 1+S\right) }\frac{(gN)^{2}}{\left( \sigma
^{2}\right) ^{3}}-\frac{1}{\left( \sigma ^{2}\right) ^{2}}+\omega _{\perp
}^{2}=0.  \label{cubic}
\end{equation}%
It is easy to see that Eq. (\ref{cubic}) has physical (positive) solutions
for $\sigma ^{2}$ if the normalized number of atoms is smaller than a
critical value,
\begin{equation}
N^{2}<N_{\mathrm{cr}}^{2}\left( \omega _{\perp }\right) \equiv \frac{\pi
^{2}(1+S)}{\sqrt{3}C_{s}^{2}g^{2}\omega _{\perp }}.  \label{sigma-max}
\end{equation}%
The existence of largest $N$ beyond which stationary solutions do not exist
reflects an obvious fact that the underlying axisymmetric 3D GP equation (%
\ref{GPE}) gives rise to collapse if $N$ is too large \cite{Carr}. Equations
(\ref{sigma-max}) and (\ref{C}) predict increase of $N_{\mathrm{cr}}$ (for
given $\omega _{\perp }$) in the lowest vortex state ($S=1$), in comparison
with its zero-vorticity counterpart, by a factor of $2\sqrt{2}$, which is a
known effect too, see Ref. \cite{DumDum} and references therein.

Alternatively, Eq. (\ref{sigma-max}) demonstrates that, for given $N$,
physical solutions exist if the transverse confinement is not too strong
(otherwise, over-squeezing of the condensate will lead to the collapse):
\begin{equation}
\omega _{\perp }<\left( \omega _{\perp }\right) _{\max }\equiv \frac{\pi
^{2}(1+S)}{\sqrt{3}\left( C_{s}gN\right) ^{2}}.  \label{omega-max}
\end{equation}%
If there is a region where $\omega _{\perp }(x)$ exceeds $\left( \omega
_{\perp }\right) _{\max }$, a moving soliton crossing into this region will
blow up due to the collapse.

Further analysis of Eq. (\ref{cubic}) demonstrates that there are two
solutions for $\sigma ^{2}$, one decreasing with $\omega _{\perp }$, and the
other one increasing. The former behavior is natural (stronger squeeze leads
to a smaller transverse size of the condensate), while the latter one is
not. To all appearance, the latter solution is unstable. Therefore, we only
take into regard the solutions with $\sigma (\omega _{\perp })$ a decreasing
function. In particular, it is easy to demonstrate that this solution
yields, for given $N$, the transverse width in the range of
\[
\frac{C_{s}|g|N}{\pi \sqrt{1+S}}\equiv \sigma _{\min }\leq \sigma <\infty ~.
\]%
Here, $\sigma _{\min }$ corresponds to $\omega _{\perp }=\left( \omega
_{\perp }\right) _{\max }$ in Eq. (\ref{omega-max}).

\section{The effective potential and equation of motion for the soliton}

If $\sigma ^{2}$ can be found from Eq. (\ref{cubic}), then $W$ must be taken
as per Eq. (\ref{W}), and both substituted in Hamiltonian (\ref{HW}). An
explicit result can be obtained in the case of a relatively weak
nonlinearity,
\begin{equation}
\omega _{\perp }(gN)^{2}\ll \frac{3\pi ^{2}\left( 1+S\right) }{C_{S}^{2}},
\label{cond}
\end{equation}%
when a relevant solution to Eq. (\ref{cubic}), calculated to first two
orders of the perturbative expansion, is
\begin{equation}
\sigma ^{2}\approx \frac{1}{\omega _{\perp }}-\frac{\left( C_{S}gN\right)
^{2}}{3\pi ^{2}\left( 1+S\right) }  \label{sigma}
\end{equation}%
(note that the first term does not depend on $S$). Then, Eq. (\ref{Hsigma})
yields the effective potential%
\begin{equation}
U_{\mathrm{eff}}(\xi )\approx N\left[ (1+S)\omega _{\perp }(\xi )+U(\xi )-%
\frac{C_{s}^{2}}{6\pi ^{2}}\left( gN\right) ^{2}\omega _{\perp }^{2}(\xi )%
\right] .  \label{approx}
\end{equation}%
For $S=0$, the appearance of combination $\omega _{\perp }(\xi )+U(\xi )$ in
potential (\ref{approx}) might be expected from the form of the last term in
Eq. (\ref{CQ}). Equation (\ref{approx}) explicitly shows how the axially
nonuniform transverse trapping induces the axial potential (which depends on
the soliton's \textquotedblleft spin"). This result may be realized as
transformation of a part of the energy of the transverse motion into the
axial potential when the confinement frequency gradually varies along $x$.

It is well known that, in the notation adopted here, an effective mass of
the soliton, if it is treated as a quasiparticle, is $M_{\mathrm{eff}}=2N$
\cite{Progress} (this relation does not depend on $S$). Thus, the equation
of motion for the soliton in effective potential (\ref{approx}), $M_{\mathrm{%
eff}}\left( d^{2}\xi /dt^{2}\right) =-\partial U_{\mathrm{eff}}/\partial \xi
$, takes the form%
\begin{equation}
\frac{d^{2}\xi }{dt^{2}}=-\frac{1}{2}\left[ \left( 1+S\right) \frac{d\omega
_{\perp }}{d\xi }+\frac{dU}{d\xi }-\frac{C_{s}^{2}}{6\pi ^{2}}\left(
gN\right) ^{2}\frac{d\left( \omega _{\perp }^{2}\right) }{d\xi }\right] .
\label{dyn}
\end{equation}%
This equation clearly shows that, in the absence of the explicit potential, $%
dU/d\xi =0$, \emph{stable} and \emph{unstable} equilibria for the soliton
are positions where the local trapping frequency attains its \emph{minimum}
and \emph{maximum}, respectively. As said above, the trapping frequency may
also be time-dependent (in addition to being $x$-dependent), $\omega _{\perp
}=\omega _{\perp }(\xi ,t)$, then the right-hand side of Eq. (\ref{dyn})
will explicitly depend on $t$.

In the most general case, without assuming condition (\ref{cond}) to hold,
Eq. (\ref{cubic}) can be solved for $\sigma ^{2}$ numerically, and the
effective potential can also be found in a numerical form. To illustrate
this possibility, in Fig. 1 we display the dependence of $\sigma ^{2}$ vs. $N
$, as obtained from the numerical solution of Eq. (\ref{cubic}) at several
fixed values of $\omega _{\perp }$, with $S=0$. As seen from the figure, the
transverse width gradually decreases with the increase of $N$, as predicted
by Eq. (\ref{sigma}). The latter effect is a natural manifestation of the
self-squeezing of the condensate with attraction between atoms.

\begin{figure}[tbp]
{\centering {{\includegraphics{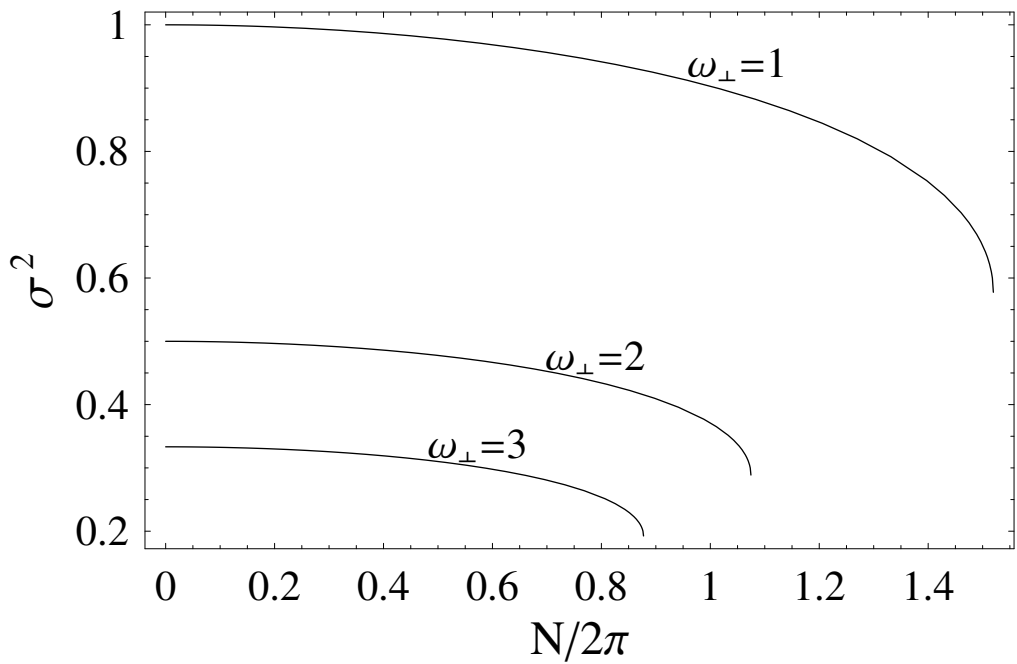}}} }
\caption{The squared transverse size of the condensate, $\protect\sigma ^{2}$%
, as a function of norm $N$, at several fixed values of the transverse
trapping frequency, $\protect\omega _{\perp }$. The curves are obtained from
numerical solution of Eq. (\protect\ref{cubic}) with $S=0$ and $g^{2}=1$,
and terminate at $N=N_{\mathrm{cr}}(\protect\omega _{\perp })$, as per Eq. (%
\protect\ref{sigma-max}).}
\label{Fig1}
\end{figure}

Results obtained by the substitution of the numerically found $\sigma ^{2}$
in Hamiltonian (\ref{Hsigma}) (with $S=0$), i.e., the effective potential,
are displayed in Fig. 2. It is seen from this figure that both the linear
dependence of the potential on $N$ for small $N$ and deviations from the
linear dependence at larger $N$ follow the analytical prediction (\ref%
{approx}).

\begin{figure}[tbp]
{\centering {{\includegraphics{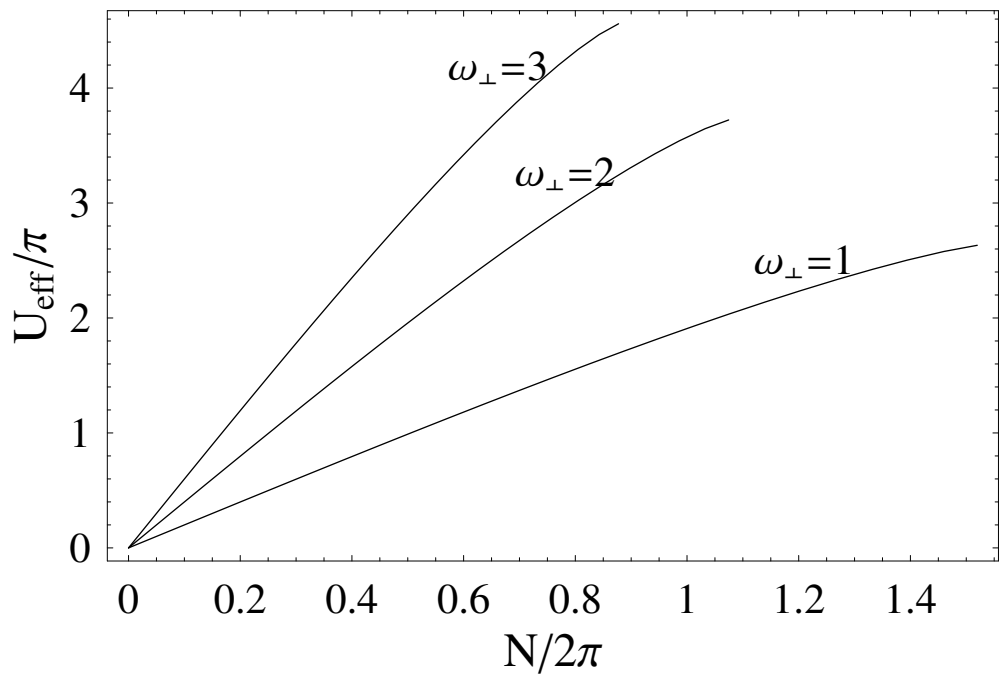}}} }
\caption{Dependences of the effective potential (\protect\ref{Hsigma}) with $%
S=0$, $U=0$ and $g^{2}=1$ on the soliton's norm $N$ at several fixed values
of $\protect\omega _{\perp },$ as obtained by the substitution of numerical
data for $\protect\sigma ^{2}$, which were used to generate Fig. 1. The
plots terminate at $N=N_{\mathrm{cr}}($ $\protect\omega _{\perp })$, the
same way as in Fig. 1.}
\label{VcNumber}
\end{figure}

\section{Conclusions}

In this work, we have demonstrated that a tight transverse trap, whose
strength gradually varies in the longitudinal (axial) direction, induces an
effective longitudinal potential for the soliton in the self-attractive BEC.
An analytical approximation for this potential, based on the variational
method, was developed, under the assumption that the longitudinal size of
the soliton is essentially smaller than the scale of the trap's
nonuniformity. A fully explicit result was obtained for the case of weak
nonlinearity. The potential depends on the soliton's intrinsic vorticity (if
any), and its norm (the number of atoms trapped in the soliton). The
analytical results reported in this Letter can be used to analyze inevitable
effects of the nonuniformity of the transverse trapping in experimental
situations, as well as to devise axially nonuniform traps with the purpose
to control the longitudinal dynamics of the solitons. Systematic numerical
simulations of the full 3D model, aimed at direct verification of the
analytical predictions, demand special effort and will be reported elsewhere.

It is relevant to mention that a similar problem can also be considered in
the model with repulsion ($g>0$ in Eq. (\ref{GPE})) and a periodic term if
the longitudinal potential generated by an optical lattice, $U_{\mathrm{OL}%
}(x)=\epsilon \cos (kx)$. In that case, the system can readily support
bright solitons of the gap type \cite{GS,HS}. However, the effective
dynamical mass of mobile gap solitons in \emph{negative} \cite{HS}; for this
reason, one may expect that, on the contrary to the ordinary solitons
considered in this Letter, gap solitons will find their stable and unstable
equilibria at points where the transverse-trapping frequency attains,
respectively, its \emph{maximum} and \emph{minimum}. These predictions are
subject to verification in numerical simulations.

B.A.M. appreciates hospitality of the CNR\ Institute of Cybernetics ``E
Caianiello" (Pozzuoli, Italy) and Department of Physical Sciences at
Universit\'{a} Federico II (Naples, Italy). The work of this author was
supported, in a part, by the Israel Science Foundation, through the
Center-of-Excellence grant No. 8006/03 (B.A.M.).

\end{document}